\documentclass[aps,prx,showpacs,superscriptaddress,twocolumn]{revtex4}

\usepackage{mathrsfs}
\usepackage{amsfonts}
\usepackage{amssymb}
\usepackage{amsmath}
\usepackage{graphicx}
\usepackage[usenames,dvipsnames]{color}
\usepackage[colorlinks=true,citecolor=blue,linkcolor=magenta]{hyperref}
\usepackage{ulem}

\newcommand{\Tr}{\mathrm{Tr}}

\newcommand{\abs}[1]{| #1 |}
\newcommand{\absLR}[1]{\left\vert #1 \right\vert}

\newcommand{\ket}[1]{\vert{ #1 }\rangle}
\newcommand{\bra}[1]{\langle{ #1 }\vert}
\newcommand{\ketbra}[2]{\vert #1 \rangle \langle #2 \vert}
\newcommand{\braket}[2]{\langle #1 \vert #2 \rangle}

\newcommand{\DeltaSO}{\Delta_\text{SO}}
\newcommand{\DeltaKK}{\Delta_{KK'}}
\newcommand{\DeltaKKh}{\Delta_{KK'\text{h}}}
\newcommand{\gs}{g_\text{s}}
\newcommand{\gorb}{g_\text{orb}}
\newcommand{\muB}{\mu_\text{B}}
\newcommand{\bfsigma}{\boldsymbol{\sigma}}
\newcommand{\nn}{\hat{\mathbf{n}}}
\newcommand{\ii}{\hat{\mathbf{i}}}
\newcommand{\jj}{\hat{\mathbf{j}}}
\newcommand{\kk}{\hat{\mathbf{k}}}
\newcommand{\geff}{\mathbf{g}_\text{eff}}
\newcommand{\Beff}{\mathbf{B}_\text{eff}}
\newcommand{\Bv}{\mathbf{B}^{V}_\mathrm{eff}}
\newcommand{\Bvv}{\mathbf{B}^{VV}_\mathrm{eff}}
\newcommand{\Heff}{H_\mathrm{eff}}
\newcommand{\sigmaK}{\bfsigma_\text{K}}
\newcommand{\JI}{J_\text{I}}

\begin{document}

\title{
Electrically driven spin resonance in a bent disordered carbon nanotube
}

\author{Ying Li}

\author{Simon C. Benjamin}

\author{G. Andrew D. Briggs}

\author{Edward A. Laird}

\affiliation{Department of Materials, University of Oxford, Parks Road, Oxford OX1 3PH, United Kingdom}

\date{\today}

\begin{abstract}
Resonant manipulation of carbon nanotube valley-spin qubits by an electric field is investigated theoretically. We develop a new analysis of electrically driven spin resonance exploiting fixed physical characteristics of the nanotube: a bend and inhomogeneous disorder. The spectrum is simulated for an electron valley-spin qubit coupled to a hole valley-spin qubit and an impurity electron spin, and features that coincide with a recent measurement are identified. We show that the same mechanism allows resonant control of the full four-dimensional spin-valley space.
\end{abstract}

\maketitle

\section{Introduction}

Gate-defined carbon nanotube quantum dots offer a clean nuclear-spin environment~\cite{Laird2014} and can be fabricated with very low disorder~\cite{Cao2005,Wu2010,Pei2012}. This makes them attractive materials for quantum devices based on electron spins~\cite{Loss1998,NielsenBook,Hanson2007}. Their strong spin-orbit coupling~\cite{Ando2000,Kuemmeth2008,Steele2013} enables qubit manipulation~\cite{Bulaev2008} by electrically driven spin resonance (EDSR). This spin-orbit mediated EDSR proceeds through several mechanisms~\cite{Laird2014}, including via a bend in the nanotube~\cite{Flensberg2010} or inhomogeneous valley mixing~\cite{Palyi2011,Szechenyi2013}, and was recently used to coherently manipulate a nanotube qubit~\cite{Laird2013}. In that experiment, the EDSR spectrum was found to be more complicated than expected from existing theories~\cite{Bulaev2008,Ando2000} taking account of the spin and valley degrees of freedom. The spectrum was found to have a substantial zero-field splitting, and to depend on gate voltage, making the qubit susceptible to electric field noise. Furthermore, although the dominant EDSR effect was concluded to result from a bend in the nanotube, the EDSR intensity did not vanish for a parallel field orientation, where the bend-mediated mechanism was predicted to be ineffective~\cite{Flensberg2010}.

Here we present a unified theory of nanotube EDSR incorporating both the effects of the bend and of inhomogeneous valley mixing. We use this to calculate the EDSR Rabi frequency as a function of magnetic field direction for a single isolated electron, whose states form a Kramers (or `valley-spin') qubit. Carefully accounting for all second-order terms in perturbation theory, we find that even for pure bend-mediated EDSR, there is an additional effect of the same order as that considered in Ref.~\cite{Flensberg2010} that gives rise to a finite EDSR frequency for parallel field, as observed experimentally~\cite{Laird2013}.

We apply our theory to the situation of two qubits in a double quantum dot. We find that realistic inter-dot tunneling strongly modifies the spectrum, giving rise to a zero-field splitting similar to that observed experimentally~\cite{Laird2013}. We further consider the effect on the spectrum of a single spin impurity coupled to one of the quantum dots~\cite{Chorley2011}. These perturbations explain some but not all of the observed deviations from previous theory.

Finally, we consider coherent manipulation of the full four-dimensional Hilbert space formed by the spin and valley degrees of freedom of a single electron, which together encode two logical qubits. We find that for realistic parameters, rapid high-fidelity operations are possible between any pair of states.

\section{Model}

\begin{figure}[tbp]
\includegraphics[width=0.8\linewidth]{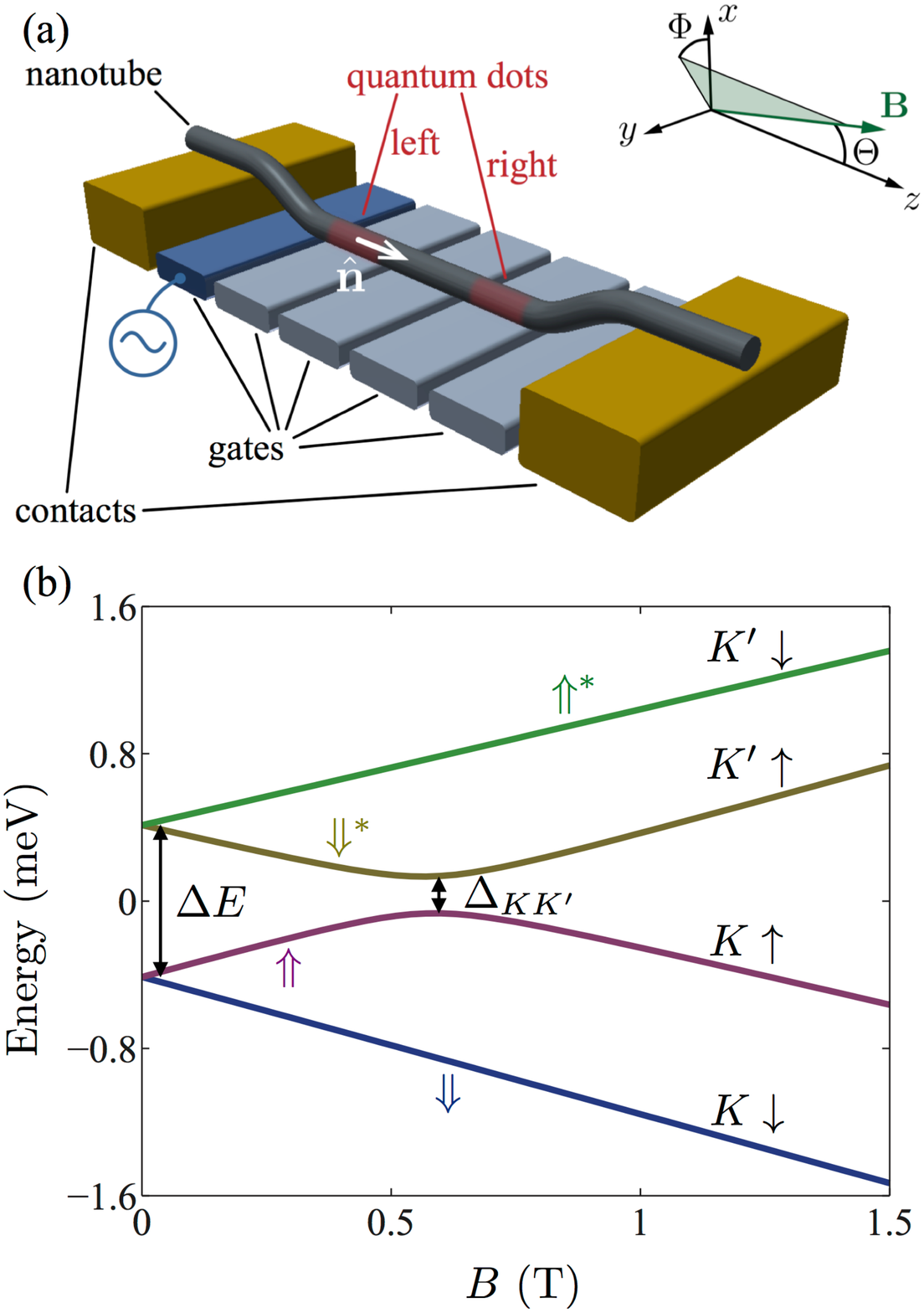}
\vspace{-0.3cm}
\caption{(a) A bent carbon nanotube device. Left and right quantum dots are formed using gate voltages to create a pair of potential wells. As indicated, one gate acts as a microwave antenna for driving EDSR transitions, which can be detected via the current between two contact electrodes. The bend is parameterized by a local tangent vector $\nn$. Inset: Coordinates used in this paper, including the polar angles $(\Theta, \Phi)$ that characterize the direction of the magnetic field~$\mathbf{B}$. (b) Spectrum of a single-electron nanotube quantum dot in a magnetic field parallel to $\nn$, taking $\DeltaSO=0.8\text{ meV}$, $\DeltaKK=0.2\text{ meV}$, $\gs=2$, and $\gorb=24$. These values are similar to those measured by transport spectroscopy~\cite{Laird2013}. The four energy eigenstates $\{\Uparrow, \Downarrow, \Uparrow^*, \Downarrow^*\}$, are labelled, as are their high-field limits $\{K\uparrow, K\downarrow, K'\downarrow, K'\uparrow\}$. The Kramers splitting $\Delta E$ and valley-mixing splitting~$\DeltaKK$ are indicated.
}
\label{setup}
\end{figure}

We begin by considering an electron confined in a single nanotube quantum dot, for example the left dot in Fig.~\ref{setup}(a)~\cite{Laird2014}. As well as its spin states $\{\uparrow, \downarrow\}$, the electron has two valley states $\{K,K'\}$ associated with opposite orbital magnetic moments along the nanotube~\cite{Minot2004}. These two degrees of freedom are coupled by spin-orbit interaction~\cite{Kuemmeth2008}, which splits states with parallel and antiparallel spin and valley magnetic moments. Additionally, the two valley states are coupled to each other, reflecting electrical disorder as well as mixing via contact electrodes.

The effect of the bend is captured by a local tangent unit vector $\nn(z)$ whose direction varies with position $z$. Likewise, the inhomogeneous valley mixing is parameterized by its position-dependent magnitude $\DeltaKK(z)$ and phase $\varphi(z)$~\cite{Palyi2011,Reynoso2012}. If the quantum dot is centered at~$z$ and its extent is much less than both the bend radius and the valley mixing correlation length, the evolution of spin and valley states in a magnetic field $\mathbf{B}$ is described by the Hamiltonian~\cite{Flensberg2010}:
\begin{eqnarray}
H(z) &=& -\frac{1}{2} \DeltaSO \tau_3 \nn(z) \cdot \bfsigma \notag \\
&& -\frac{1}{2} \DeltaKK(z) [\tau_1\cos\varphi(z)+\tau_2\sin\varphi(z)] \notag \\
&& +\frac{1}{2} \gs \muB \mathbf{B} \cdot \bfsigma
+\frac{1}{2} \gorb \muB \mathbf{B} \cdot \nn(z) \tau_3.
\label{eq:Hz}
\end{eqnarray}
Here $\tau_{i=1,2,3}$ and $\sigma_{i=x,y,z}$ are respectively the Pauli operators acting in valley and spin space, $\DeltaSO$ is the spin-orbit coupling, and the spin and orbital $g$-factors are denoted respectively $\gs$ and $\gorb$. The coordinates $\{x,y,z\}$, associated with unit vectors $\{\ii, \jj, \kk\}$, are defined in Fig.~\ref{setup}(a).

At $\mathbf{B}=0$, the four valley-spin states form two Kramers doublets separated by an energy gap $\Delta E(z) = \sqrt{\DeltaSO^2+\DeltaKK(z)^2}$. Each doublet is an effective spin-1/2 system whose states are denoted $\{\ket{\Uparrow}, \ket{\Downarrow}\}$ for the lower doublet and $\{\ket{\Uparrow^*}, \ket{\Downarrow^*}\}$ for the upper doublet. Either doublet can be operated as a valley-spin qubit~\cite{Flensberg2010,Laird2013}.

Applying a magnetic field splits these effective spin states~[Fig.~\ref{setup}(b)]. This is equivalent to a Zeeman splitting, but with an effective gyromagnetic tensor $\geff(z)$ that is anisotropic (and therefore position-dependent) because of the axial magnetic moment associated with the valley degree of freedom~\cite{Minot2004}. Alternatively, the anisotropy can be described by an effective magnetic field $\Beff(z) \equiv \geff(z) \cdot \mathbf{B} / \gs$ that is tilted and scaled relative to $\mathbf{B}$. As shown in the next section, the effective Zeeman Hamiltonian within each doublet is~\cite{Flensberg2010}:
\begin{eqnarray}
H_\text{K}(z)
&=& \frac{1}{2} \muB \mathbf{B} \cdot \geff(z) \cdot \sigmaK \notag \\
&=& \frac{1}{2} \gs \muB \Beff(z) \cdot \sigmaK,
\label{HK}
\end{eqnarray}
where $\sigmaK$ are Pauli operators of the effective spin-1/2 (Kramers qubit) with states $\ket{\Uparrow}$ (or $\ket{\Downarrow^*}$) having eigenvalue $+1$ and $\ket{\Downarrow}$ (or $\ket{\Uparrow^*}$) having eigenvalue $-1$.

\section{EDSR of a single qubit}
\label{EDSR}

In this section, we calculate the Rabi frequency for a qubit defined in a single quantum dot, accounting for both the bend and inhomogeneous disorder. We go beyond previous work \cite{Flensberg2010} by fully incorporating these effects up to second order in perturbation theory. The valley-spin qubit can be manipulated by applying a microwave gate voltage to oscillate the quantum dot position along the nanotube, as recently proposed~\cite{Flensberg2010} and demonstrated~\cite{Pei2012,Laird2013}. This effect is understood as a result of the inhomogeneous field $\Beff(z)$ experienced by the electron, which coherently drives resonant transitions within the doublet~\cite{Flensberg2010}. We neglect the comparatively weak Rashba-like spin-orbit coupling to the electric field~\cite{Bulaev2008,Klinovaja2011}.

To investigate the evolution of the driven qubit, we perform a perturbation calculation based on the Hamiltonian of Eq.~(\ref{eq:Hz}), with the aim of deriving an effective Hamiltonian governing transitions between qubit states. The perturbation parameters are taken as $\mathbf{B}$ and the displacement $\delta z = z-z_0$ of the quantum dot, where~$z$ is the instantaneous dot center position and $z_0$ the position without driving. Axes are chosen so that at $z_0$ the nanotube is aligned with the $z$-axis and bent in the $x$-$z$ plane~[Fig.~\ref{setup}(a)]. This gives $\nn(z) = \cos\theta(z)\kk + \sin\theta(z)\ii$, where $\theta(z)$ is the angle between the nanotube and the $z$-axis. Without loss of generality, the valley-mixing phase is defined so that $\varphi(z_0)=0$. We define the unperturbed Hamiltonian $H_0 \equiv H (z_0)|_{\mathbf{B} = 0}$ and the perturbation $V \equiv H(z) - H_0$, and truncate at second order in the perturbation parameters:
\begin{equation}
V \simeq V^{(1)}+V^{(2)}.
\end{equation}
In turn, the first-order perturbation $V^{(1)} \equiv V_z+V_B$ is a sum of terms proportional respectively to $\delta z$ and to~$\mathbf{B}$:
\begin{equation}
V_z = -\frac{\delta\theta}{2} \DeltaSO \tau_3 \sigma_x
-\frac{\delta_{KK'}}{2} \DeltaKK \tau_1 -\frac{\delta\varphi}{2} \DeltaKK \tau_2,
\label{eq:Vz}
\end{equation}
and
\begin{equation}
V_B = \frac{1}{2} \gs \muB \mathbf{B} \cdot \bfsigma
+\gorb \muB B_z \tau_3.
\label{eq:VB}
\end{equation}
The second order perturbation is:
\begin{eqnarray}
V^{(2)} &=& \frac{\delta\theta^2}{4} \DeltaSO \tau_3 \sigma_z
+\frac{\delta\varphi}{2} \DeltaKK \left( \frac{\delta\varphi}{2} \tau_1 - \delta_{KK'} \tau_2 \right) \notag \\
&&+ \gorb \muB \delta \theta B_x \tau_3.
\label{eq:V2}
\end{eqnarray}
Here, we define $\delta\theta \equiv \delta z \partial_z \theta |_{z=z_0}$, $\delta_{KK'} \equiv \delta z (\DeltaKK^{-1} \partial_z \DeltaKK) |_{z=z_0}$, and $\delta\varphi \equiv \delta z \partial_z \varphi |_{z=z_0}$.

From these perturbations an effective Hamiltonian can be derived in the subspace of a single doublet. To simplify this derivation, we define an energy quantum number which takes the value $g$ ($e$) for the lower (upper) Kramers doublet and a Kramers quantum number that takes the value 0 or 1 to label the state in each doublet. In this notation, the four valley-spin states at zero field [Fig.~\ref{setup}(b)] are written $\ket{\Uparrow}=\ket{g 0}$, $\ket{\Downarrow}=\ket{g 1}$, $\ket{\Uparrow^*}=\ket{e 1}$, and $\ket{\Downarrow^*}=\ket{e 0}$ (see Appendix~\ref{AppendixA}). The unperturbed Hamiltonian is $H_0 = -(\Delta E_0/2) \sigma_\text{E}^z$, where $\Delta E_0 \equiv \Delta E(z_0)$, $\sigma_\text{E}^z$ is the Pauli operator in the energy subspace $\{\ket{g},\ket{e}\}$.

The effective Hamiltonian can now be derived to second order in $B$ and $\delta z$. Writing 
\begin{equation}
H_{\text{eff}} = H_{\text{eff}}^{(1)} + H_{\text{eff}}^{(2)},
\label{eq:Heff}
\end{equation}
the first and second order terms are~\cite{Kitaev2006}:
\begin{equation}
H_{\text{eff}}^{(1)} = P_g V^{(1)} P_g + P_e V^{(1)} P_e
\label{eq:Heff1}
\end{equation}
and
\begin{eqnarray}
H_{\text{eff}}^{(2)} &=& P_g V^{(2)} P_g + P_e V^{(2)} P_e \notag \\
&&+ P_g V^{(1)} G'_g V^{(1)} P_g + P_e V^{(1)} G'_e V^{(1)} P_e.
\label{eq:Heff2}
\end{eqnarray}
Here $P_g \equiv (1+\sigma_\text{E}^z)/2$ and $P_e \equiv (1-\sigma_\text{E}^z)/2$ are respectively projectors onto the subspaces of the upper and lower doublets, and $G'_g \equiv P_e (-\Delta E_0/2-H_0)^{-1} P_e$ and $G'_e \equiv P_g (\Delta E_0/2-H_0)^{-1} P_g$ are Green's functions. This perturbation theory is valid for, $\delta\theta\DeltaSO$, $\delta_{KK'}\DeltaKK$, $\delta\varphi\DeltaKK$, $\gs\muB\abs{\mathbf{B}}$, $\gorb\muB B_z \ll \Delta E_0$.

\begin{figure}[tbp]
\includegraphics[width=1\linewidth]{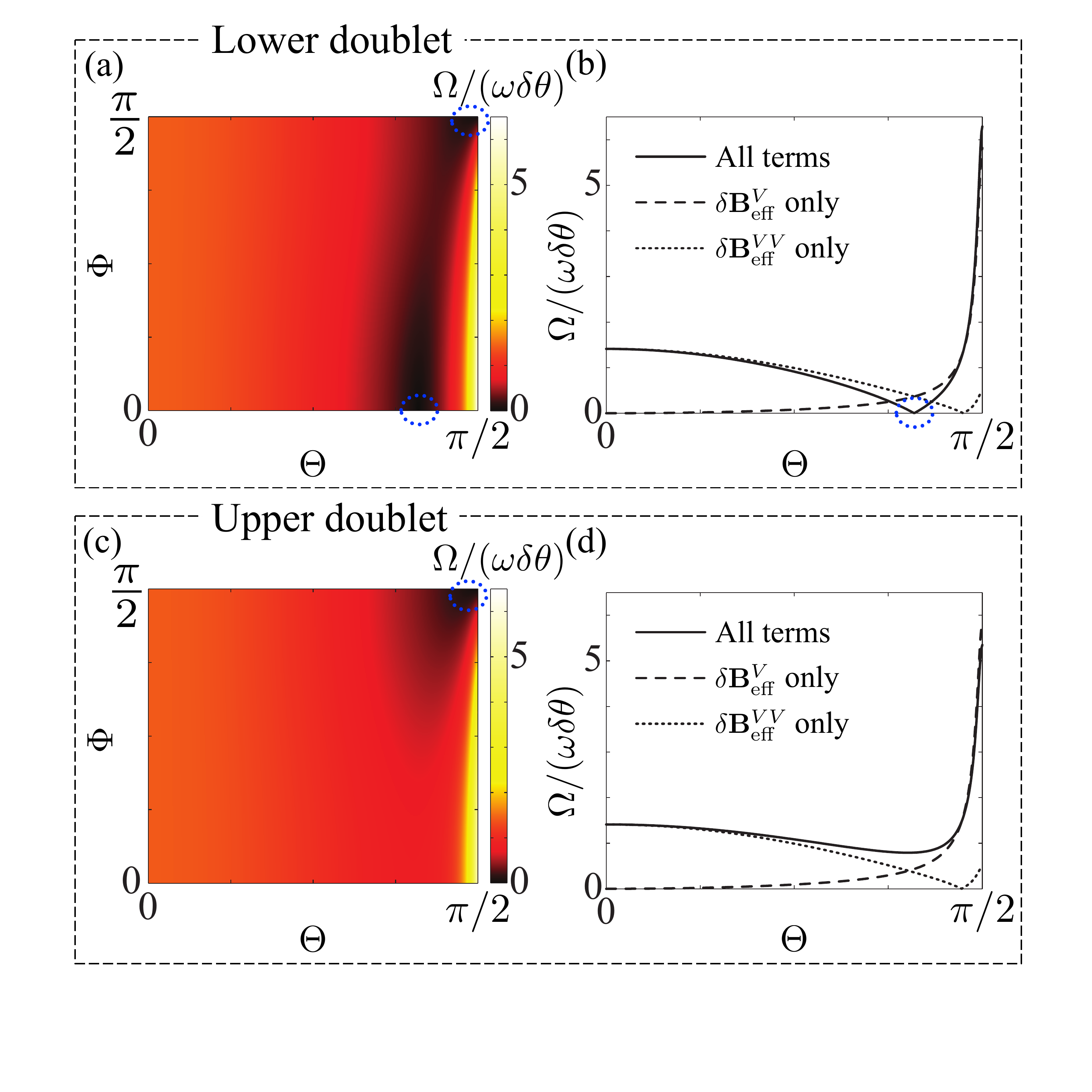}
\caption{
Rabi frequency in a bent nanotube with homogeneous valley mixing ($\DeltaKK$ and $\varphi$ independent of $z$) as a function of magnetic field angles [defined in Fig.~\ref{setup}(a)]. (a) Rabi frequency $\Omega$ in the lower Kramers doublet, calculated according to Eq.~(\ref{Omega}) and normalized by the free-electron Larmor frequency $\hbar\omega = \gs\muB\abs{\mathbf{B}}$ and the driving amplitude $\delta\theta$. (b) Rabi frequency as a function of $\Theta$ for $\Phi=0$, calculated by Eq.~(\ref{Omega}) (solid). Separate contributions from $\delta\Bv$~\cite{Flensberg2010} and $\delta\Bvv$ are marked with dashed and dotted lines respectively. Dashed ellipses mark field angles where $\Omega=0$. (c-d) The same plots for the upper doublet. Whereas for the lower doublet $\Omega$ vanishes at two field angles, for the upper doublet it vanishes only for $\mathbf{B}$ along $y$. Throughout this figure, we take $\DeltaSO/\DeltaKK=4$ and $\gorb/\gs=12$, consistent with~Fig.~\ref{setup}.
}
\label{RFbend}
\end{figure}

To first order, the first term of Eq.~(\ref{eq:Vz}) (due to spin-orbit coupling) couples only states in different doublets, while the second and third terms (due to valley mixing) shift the energy of an entire doublet (see Appendix~\ref{AppendixB}). The first order part of the effective Hamiltonian is therefore [by Eq.~(\ref{eq:Heff1})]:
\begin{equation}
H_\text{eff}^{(1)} = H_\text{K}(z_0) + C_1\times \sigma_\text{E}^z,
\end{equation}
where the effective magnetic field is
\begin{eqnarray}
\mathbf{B}_{\text{eff}} &=& \sin \chi B_x \ii +\sin \chi B_y \jj \notag \\
&&+ [1 + \sigma_\text{E}^z (\gorb/\gs) \cos \chi] B_z \kk,
\end{eqnarray}
we define $\cos \chi = \DeltaSO/\Delta E_0$ and $\sin \chi = \DeltaKK(z_0)/\Delta E_0$, and $C_1$ is a scalar. Comparison with Eq.~(\ref{HK}) shows that the effective gyromagnetic tensor is:
\begin{equation}
\mathbf{g}_{\text{eff}} = \gs
\left( \begin{array}{ccc}
\sin \chi & 0 & 0 \\
0 & \sin \chi & 0 \\
0 & 0 & 1 + \sigma_\text{E}^z (\gorb/\gs) \cos \chi
\end{array} \right).
\label{geff}
\end{equation}
The parallel component of $\geff$ differs between doublets because of the different relative alignment of spin and valley magnetic moments.

EDSR arises from the second-order effective Hamiltonian, $H^{(2)}_\mathrm{eff}$. The existence of bend-mediated EDSR can already be seen by evaluating the first two terms of~Eq.~(\ref{eq:Heff2}):
\begin{multline}
P_g V^{(2)} P_g + P_e V^{(2)} P_e \\
= \frac{1}{2} \gs \muB \delta \Bv(z) \cdot \sigmaK
+ C_V(z) \times \sigma_\text{E}^z,
\label{eq:HV}
\end{multline}
where
\begin{eqnarray}
\delta\Bv = \delta\theta (\sigma_\text{E}^z \gorb \cos \chi / \gs) B_x \kk,
\label{deltaB1}
\end{eqnarray}
is the $z$-dependent effective magnetic field and $C_V(z)$ is a scalar function. Under microwave driving, the position of the quantum dot oscillates sinusoidally, $\delta z = \delta z_\text{A} \sin (2 \pi f t)$ with amplitude $\delta z_\text{A}$, giving rise to resonant transitions between states in the same doublet via the time-varying $\delta \Bv$. This was identified already in the first theory of bend-mediated EDSR~\cite{Flensberg2010}. From Eq.~(\ref{deltaB1}), it is clear that this EDSR effect vanishes if $\mathbf{B}$ is applied in the $y$-$z$ plane.

\begin{figure}[tbp]
\includegraphics[width=1\linewidth]{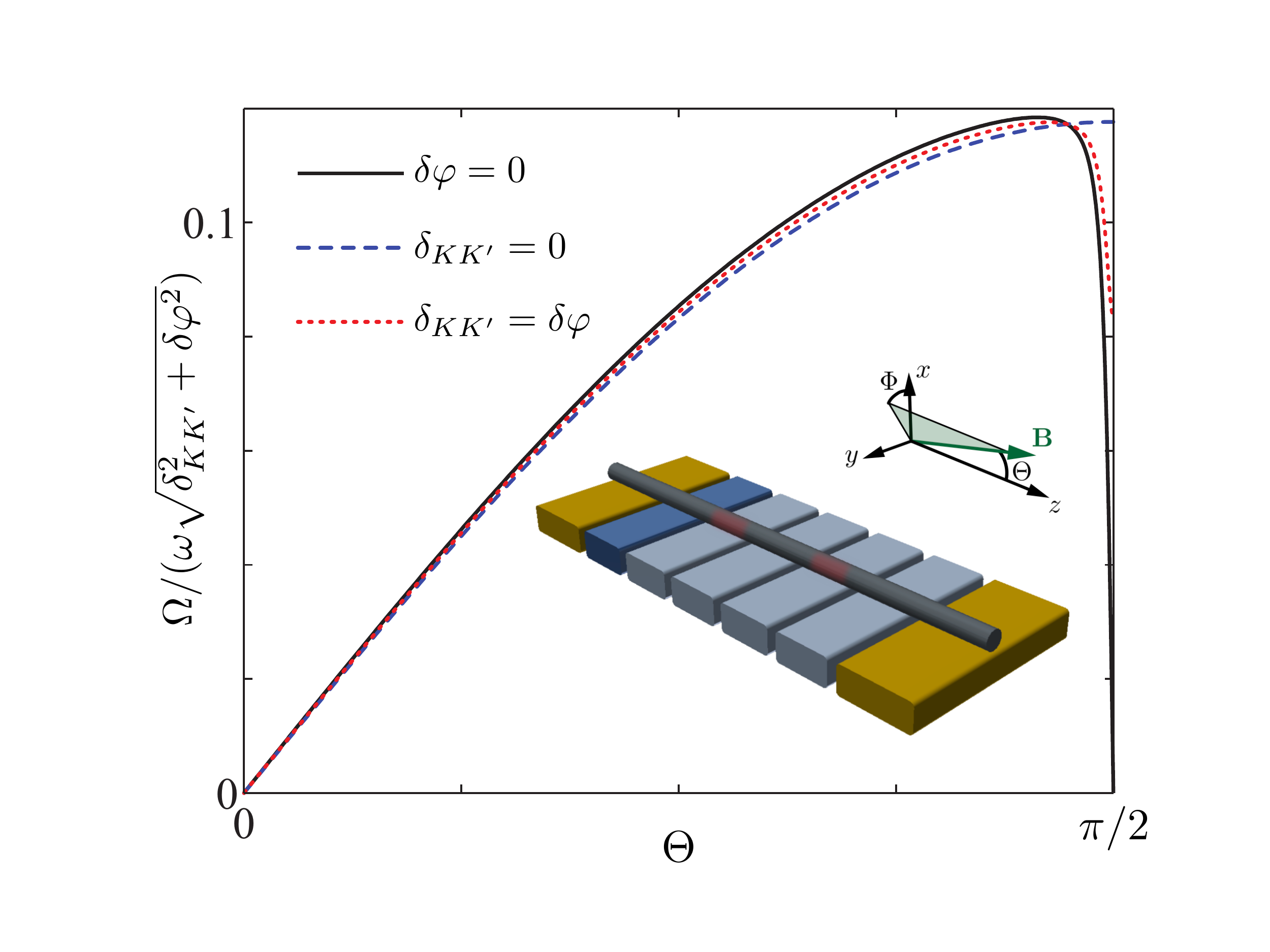}
\caption{
Rabi frequency in a straight nanotube device ($\Theta=0$, see inset) with inhomogeneous valley mixing, calculated according to Eq.~(\ref{Omega}) as a function of $\Theta$. The Rabi frequency is the same for both doublets. By symmetry, $\Omega$ is independent of $\Phi$. Three cases are considered: inhomogeneous valley-mixing amplitude only ($\delta\varphi=0$), inhomogeneous valley-mixing phase only ($\delta_{KK'}=0$), and both effects combined ($\delta_{KK'}=\delta\varphi$). For comparison, $\Omega$ is normalized by $\sqrt{\delta_{KK'}^2+\delta\varphi^2}$ and $\hbar\omega = \gs\muB\abs{\mathbf{B}}$. This figure assumes the same numerical parameters as Fig.~\ref{RFbend}.
}
\label{RFstraight}
\end{figure}

For a full theory of EDSR, it is necessary to include the last two terms of Eq.~(\ref{eq:Heff2}), which contribute in the same order in $\mathbf{B}$ and $\delta z$. Their contribution to the effective Hamiltonian is (see Appendix~\ref{AppendixB})
\begin{multline}
 P_g V^{(1)} G'_g V^{(1)} P_g + P_e V^{(1)} G'_e V^{(1)} P_e\\
= \frac{1}{2} \gs \muB \delta\Bvv(z) \cdot \sigmaK
+ C_{VV}(z) \times \sigma_\text{E}^z,
\end{multline}
where as in Eq.~(\ref{eq:HV}) the coupling is equivalent to an effective magnetic field, in this case:
\begin{eqnarray}
&& \delta \Bvv = \frac{1}{2} \{ \notag \\
&& \begin{array}{r c l}
\times[ & \sigma_\text{E}^z \delta\theta (\gorb/\gs) \sin 2\chi B_z & ~ \\
~ & +\delta_{KK'} \sin 2\chi \cos \chi B_x - \sigma_\text{E}^z \delta\varphi \sin 2\chi B_y & ]\ii \\
+[ & \delta_{KK'} \sin 2\chi \cos \chi B_y + \sigma_\text{E}^z \delta\varphi \sin 2\chi B_x & ]\jj \\
+[ & \delta\theta (1+\cos 2\chi) B_x & ~ \\
~ & - \sigma_\text{E}^z \delta_{KK'} (\gorb/\gs) \sin \chi \sin 2\chi B_z & ]\kk\}.
\end{array}
\label{B2}
\end{eqnarray}
Here, $C_{VV}(z)$ is a scalar function. Finally, substitution into Eq.~(\ref{eq:Heff}) gives for the effective Hamiltonian in each Kramers doublet
\begin{equation}
\Heff= \frac{1}{2} \muB \gs [\Beff + \delta \Beff (z)] \cdot \sigmaK,
\label{eq:HeffBeff}
\end{equation}
where the inhomogeneous effective magnetic field is
\begin{eqnarray}
\delta\Beff = \delta\Bv + \delta\Bvv,
\label{eq:deltaB}
\end{eqnarray}
and a term proportional to $\sigma_\text{E}^z$ has been neglected.

This analysis shows that the two contributions $\delta\Bv$ and $\delta\Bvv$ to the inhomogeneous effective magnetic field lead to two different mechanisms of EDSR. When the dot position oscillates, the orbital magnetic moment experiences an \textit{external} inhomogeneous magnetic field [see last term in Eq.~(\ref{eq:V2})], which couples two states in the same doublet and results in the contribution $\delta\Bv$. The external magnetic field experienced by the spin is homogeneous because the spin magnetic moment is decoupled from the dot position. However, the spin experiences an \textit{internal} inhomogeneous magnetic field, which is due to spin-orbit coupling and depends on the valley state [see first term in Eq.~(\ref{eq:Vz})]. As the quantum dot moves along a bend, the orientation of the orbital magnetic moment changes, resulting in a spin flip due to spin-orbit coupling. In an magnetic field, this spin flip can lead to a transition within the doublet. This internal inhomogeneous magnetic field results in terms proportional to $\delta\theta$ in $\delta\Bvv$. These two mechanisms contribute comparably to EDSR of a valley-spin qubit. The external-field mechanism is stronger for $\mathbf{B}$ perpendicular to the nanotube, while the internal-field mechanism is stronger for $\mathbf{B}$ parallel. As a result, the valley-spin qubit can be manipulated in parallel magnetic field, as observed recently~\cite{Pei2012,Laird2013}. Moreover, if disorder is inhomogeneous, valley states may be flipped due to varying valley mixing, allowing the valley-spin qubit to be operated in a straight nanotube.

The Rabi frequency $\Omega$ of EDSR transitions is proportional to the component of $\delta \Beff$ perpendicular to $\Beff$:
\begin{equation}
\hbar\Omega = \frac{1}{2} \gs \muB \abs{\delta\Beff^\perp}_{z=z_0+\delta z_\text{A}},
\label{Omega}
\end{equation}
where
\begin{equation}
\delta\Beff^\perp = \delta\Beff - \frac{\delta\Beff\cdot\Beff}{{\abs{\Beff}}^{2}} \Beff.
\end{equation}
To distinguish the separate EDSR effects, Fig.~\ref{RFbend} shows the Rabi frequency within both doublets as a function of field angle for a nanotube with homogeneous disorder, so that only the bend-mediated effect is active. Because~$\geff$ differs between the two doublets [Eq.~(\ref{geff})], $\Omega$ also differs, as seen by comparing Fig.~\ref{RFbend}(a-b) and (c-d). For both doublets, $\Omega$ vanishes when $\mathbf{B}$ is directed along the $y$-axis, because $\delta \Beff$ is then parallel to $\Beff$. Additionally, for positive $\DeltaSO$, interference between the two terms in~Eq.~(\ref{eq:deltaB}) leads to vanishing $\Omega$ in the lower doublet at one field angle in the $x-z$ plane. For negative $\DeltaSO$ a similar vanishing point occurs in the upper doublet. This is clear from plots of $\Omega$ as a function of field angle in the plane of the bend [Fig.~\ref{RFbend}(b,c)]. Including $\delta \Bvv$ as well as $\delta \Bv$ in the analysis shows that the bend mechanism alone leads to non-zero $\Omega$ at parallel $\mathbf{B}$, consistent with experiment~\cite{Laird2013}.

Figure~\ref{RFstraight} shows the angle dependence of $\Omega$ for a straight nanotube. This differs depending whether EDSR is mediated by inhomogeneous magnitude $\DeltaKK$ or inhomogeneous phase $\varphi$ of the valley mixing parameter; the difference becomes especially marked for perpendicular field, where $\Omega$ vanishes for inhomogeneous phase but is maximal for inhomogeneous magnitude.

\section{EDSR spectrum of coupled qubits}

\begin{figure}[tbp]
\includegraphics[width=1\linewidth]{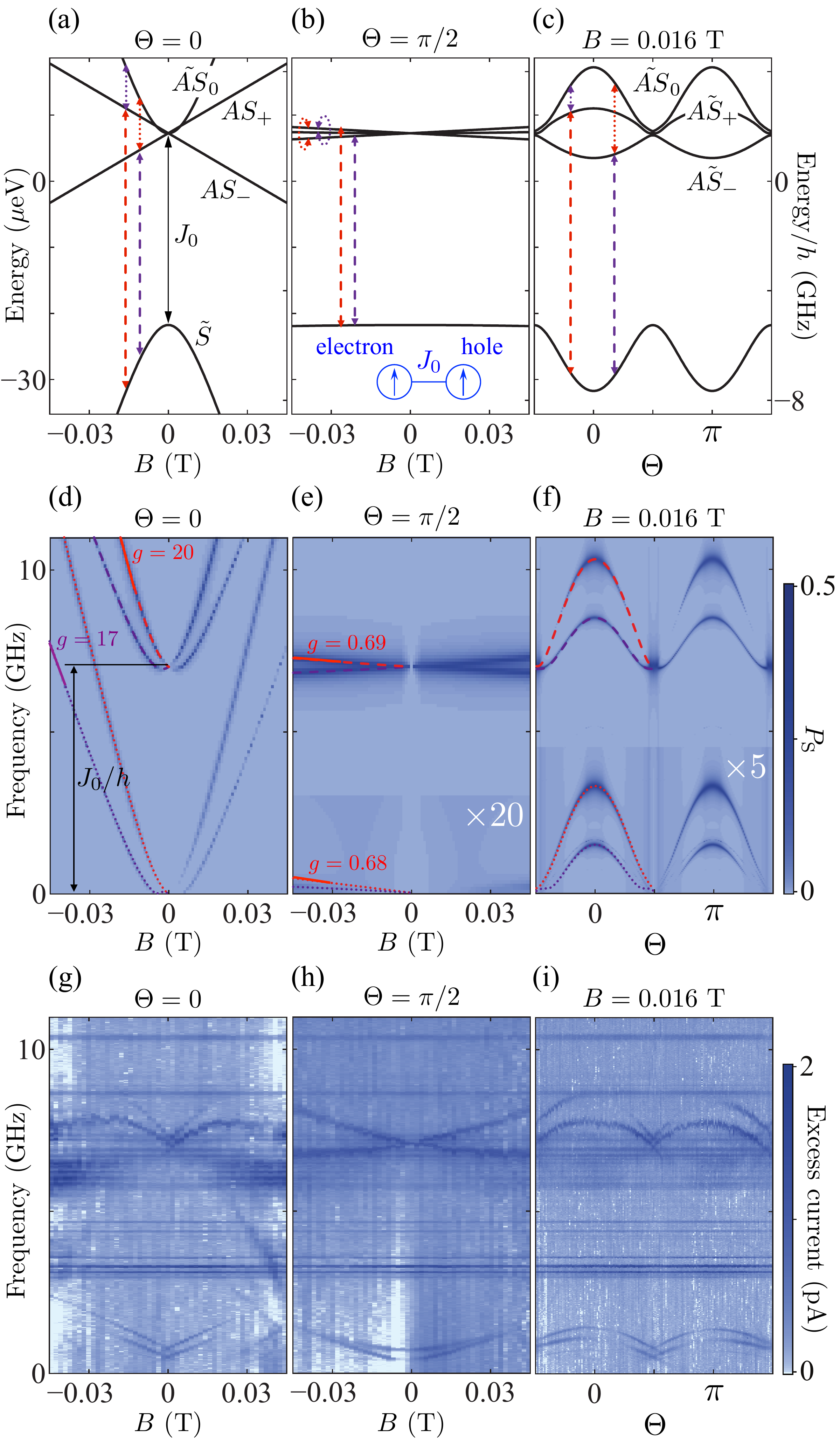}
\caption{ Energy levels and EDSR spectra of two coupled valley-spin qubits. (a--c) Simulated energy levels for parallel (a) and perpendicular (b) magnetic field, and of field angle at constant magnitude (c). For all plots, the field is in the $x$-$z$ plane, i.e., $\Phi=0$. States are labelled $\tilde{S}$, $\tilde{AS}$ according to their longitudinal symmetry at $B=0$ (see text). Inset to~(b) shows schematically the quantum dot arrangement. (d--f) Simulated EDSR transition spectra for the same conditions. Selected transitions (marked with arrows in the energy spectra) are highlighted. Corresponding $g$-factors obtained from the line slopes are marked. In (e--f), $P_\mathrm{S}$ is scaled in parts of the plots to make the low-frequency resonances clearer. (g--i): Experimental EDSR transport spectra measured in~\cite{Laird2013}. Simulations assume $\DeltaSO=0.8\text{ meV}$, $\gorb=24$ for both quantum dots, $\DeltaKK=0.2\text{ meV}$ and $\varphi=0$ for the electron quantum dot, and valley-mixing strength $\DeltaKKh=0.6\text{ meV}$ and phase $\varphi_\text{h}=0.005\pi$ for the hole quantum dot. Values for $\DeltaSO$, $\gorb$, and $\DeltaKK$ correspond to values deduced from d.c.\ transport in~\cite{Laird2013}. Values for $\DeltaKKh$ and $\varphi_\text{h}$ are chosen to approximately match EDSR spectra in the same device, requiring $\DeltaKKh$ to be slightly higher than indicated by d.c.\ transport. Exchange interaction is set to $J_0/h = 7\text{ GHz}$ to match the observed zero-field splitting.}
\label{spectrum2dots}
\end{figure}

This section presents numerical calculations of the EDSR spectrum according to the above model. As in recent experiments, we assume the spectrum is measured by exploiting Pauli blockade in a double quantum dot, where spin and valley flips lead to an enhancement of the leakage current~\cite{Ono2002,Hanson2007,Pei2012,Laird2013}. We go beyond previous work~\cite{Flensberg2010,Szechenyi2013,Osika2014} by including the effect of tunnel coupling between quantum dots \cite{Nowak2011}.

\subsection{Two coupled quantum dots}

We consider a nanotube double quantum dot configured with a single electron in the left dot and a single hole in the right~\cite{Laird2013}. Taking the inter-doublet gap $\Delta E$ as large compared to temperature and source-drain bias, only the first doublet in each dot need be considered, so that the electron and the hole each form a valley-spin qubit with Pauli operators $\{\sigma_\text{e}^{i=x,y,z}\}$ and $\{\sigma_\text{h}^{i=x,y,z}\}$ respectively. The total Hamiltonian of the system is:
\begin{equation}
H_\text{DD} = H_\text{e} + H_\text{h} + H_\text{eh},
\label{eq:Htotal}
\end{equation}
where $H_\text{e}$ is the Hamiltonian of an isolated electron qubit, $H_\text{h}$ is the Hamiltonian of an isolated hole qubit, and $H_\text{eh}$ describes the coupling between them. Both $H_\text{e}$ and $H_\text{h}$ are of the form of Eq.~(\ref{eq:HeffBeff}), with the electron Hamiltonian including the effect of driving:
\begin{equation}
H_\text{e} = \frac{1}{2} \muB \gs [\Beff^\text{e} + \delta \Beff^\text{e} (z) ] \cdot \bfsigma_\text{e}
\label{eq:He}
\end{equation}
and
\begin{equation}
H_\text{h} = \frac{1}{2} \muB \gs \Beff^\text{h} \cdot \bfsigma_\text{h}.
\end{equation}
Here the $\sigma_\mathrm{e}^z=\{1,-1\}$ eigenstates of the electron qubit are $\{\Uparrow, \Downarrow\}$ as defined in Fig.~\ref{setup}(a), while the $\sigma_\mathrm{h}^z=\{1,-1\}$ eigenstates of the hole qubit are states with the $\{\Uparrow^*, \Downarrow^*\}$ electron levels unfilled. Effective magnetic fields for the two qubits are determined by the common spin-orbit coupling strength $\DeltaSO$ and orbital $g$-factor $g_\mathrm{orb}$, and separate valley mixing parameters for the electron qubit ($\DeltaKK$, $\varphi$) and the hole qubit ($\DeltaKKh$, $\varphi_\text{h}$). Coupling between qubits arises because of interdot tunneling. As shown in Appendix~\ref{AppendixC}, this gives an exchange-like Hamiltonian:
\begin{eqnarray}
H_\text{eh} &=& \frac{J_0}{4} [\sigma_\text{e}^z\sigma_\text{h}^z + \sigma_\text{e}^x (\sigma_\text{h}^x\cos\alpha+\sigma_\text{h}^y\sin\alpha) \notag \\
&&+ \sigma_\text{e}^y (\sigma_\text{h}^y\cos\alpha-\sigma_\text{h}^x\sin\alpha)],
\label{H2dots}
\end{eqnarray}
where $\alpha$ is determined by the relative valley mixing phase of the quantum dots and $J_0$ is the exchange strength.

The energy levels of this system are plotted in Fig.~\ref{spectrum2dots} (a-c) as a function of parallel and perpendicular magnetic field, and of field angle. At $\mathbf{B}=0$ the four two-qubit states are split by $J_0$ according to the longitudinal symmetry of the wavefunction; the three longitudinally antisymmetric states (denoted $AS_-$, $AS_0$ and $AS_+$) are raised in energy compared to the longitudinally symmetric state $S$~\cite{Laird2014}. This is analogous to the singlet-triplet splitting in conventional two-spin systems. Application of a magnetic field splits the levels further through coupling to the spin and valley magnetic moments. The $AS_{+}$ and $AS_{-}$ states are composed of qubit states with parallel magnetic moments, and therefore separate linearly in energy with parallel magnetic field. However, the $S$ and $AS_0$ states are combinations of qubit states with antiparallel magnetic moments, and are therefore mixed at finite field, forming energy eigenstates $\tilde{S}$ and $\tilde{AS}_0$. This is analogous to the mixing of $S$ and $T_0$ states by a field gradient in conventional semiconductors~\cite{Jouravlev2006}. For other field directions, $AS_{+}$ and $AS_{-}$ states are also mixed with~$S$, forming eigenstates $\tilde{AS}_{+}$ and $\tilde{AS}_{-}$. Pauli blockade applies to the $AS$ components of each eigenstate, but not to the $S$ component.

In the Pauli blockade detection scheme~\cite{Koppens2006,Laird2013}, the leakage current depends on the EDSR transition rate from blocked to unblocked states. We simulate the transition process from an initial state described by the density matrix:
\begin{eqnarray}
\rho_B = N^{-1} \sum_n e^{-\tau\Tr(\rho_S\ketbra{\psi_n}{\psi_n})} \ketbra{\psi_n}{\psi_n},
\end{eqnarray}
where $\{ \psi_n \}$ are the energy eigenstates, $\rho_S$ is the density matrix corresponding to a pure $S$ state, $\tau$ is a parameter describing how faithfully the system is prepared in a blocked state, and $N$ is a normalization factor. This is the state prepared assuming that an electron and a hole in a completely mixed state are loaded from the leads, followed by decay of the non-blocked components through their overlap with the S state. We set $\tau=10^3$, implying efficient initialization to a blocked state.

\begin{figure}[tbp]
\includegraphics[width=1\linewidth]{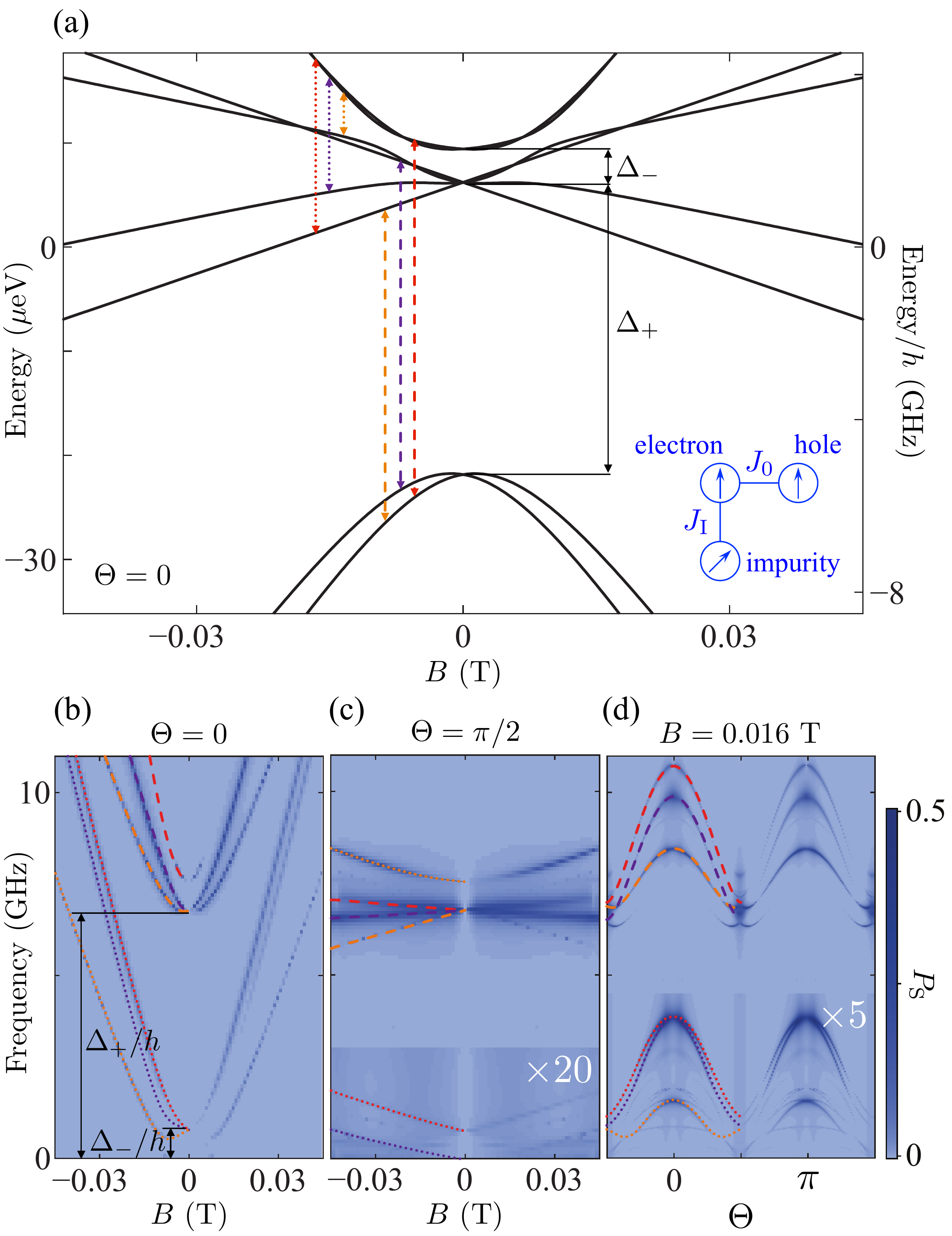}
\caption{
Energy spectrum (a) and EDSR transition spectra (b)--(d) of coupled valley-spin qubits and an impurity spin (schematic in inset of (a)). We take for the impurity $g$-factor $g_\mathrm{I}=2$ and coupling strength $J_\mathrm{I}=-1$~GHz. Other parameters are as in Fig.~\ref{RFstraight}. Magnetic field is parallel with the nanotube in (a) and (b), perpendicular in (c), and has magnitude $\abs{\mathbf{B}}=0.016\text{ T}$ in (d). Zero-field energy gaps $\Delta_+/h = 6.8\text{ GHz}$ and $\Delta_-/h = 0.8\text{ GHz}$, are indicated. Energy spectra corresponding to (c) and (d) are shown in Fig.~\ref{spectrumApp}.
}
\label{spectrumdefect}
\end{figure}

From this initial state, EDSR is simulated as a coherent time evolution governed by Eq.~(\ref{H2dots}). The right quantum dot is assumed stationary, while the left is driven according to $z = z_0 + \delta z_\text{A} \sin (2\pi f t)$, so that the time dependence enters Eq.~(\ref{H2dots}) via $H_\text{e}$ [Eq.~(\ref{eq:He})]. Valley mixing is assumed to be inhomogeneous ($\delta_{KK'}=\delta\varphi=0$). For each value of $\mathbf{B}$ in the simulation the driving amplitude is set so that $\Omega=0.1$~GHz, except where this would require $\delta \theta>0.2$, in which case the amplitude is reduced to bring $\delta\theta$ down to this value. This allows the simulation to stay within the regime of perturbation theory.

Assuming that conversion from blocked to unblocked states is the transport-limiting step, the current through the device is proportional to the averaged probability to evolve from $\rho_B$ to $\rho_S$. To obtain EDSR spectra, we calculate the average probability over a burst, $P_{\text{S}}=T^{-1}\int_0^T dt \Tr[\rho(t)\rho_{\text{S}}]$, where $\rho(t)$ is the state during a burst, $\rho_{\text{S}}$ the density matrix for the longitudinally symmetric state, and $T$ is the burst duration. We choose $T = 100\text{ ns}$, typical for spectroscopy measurements~\cite{Laird2013}.

This probability is shown in Fig.~\ref{spectrum2dots}(d-f). The resulting spectrum is more complicated than for isolated qubits. Several resonances are observed, split into two manifolds. The upper manifold (corresponding to $\tilde{S}\leftrightarrow\tilde{AS}$ transitions), is more intense than the lower manifold (corresponding to transitions within the $\tilde{AS}$ subspace) because of the greater overlap of $\tilde{S}$ with $S$. The exchange coupling $J_0$ is evident as a zero-field splitting for the upper manifold. Both manifolds are in turn split into two lines due to the different effective $g$-factors of the electron and hole qubits. Because of the procedure for choosing the driving strength described above, line intensities at different magnetic fields should not be compared.

These simulations can be compared with experimental spectra reproduced in Fig.~\ref{spectrum2dots}(g--i). (For details, see~\cite{Laird2013}. The excess current, with non-resonant background subtracted, should be interpreted as proportional to $P_\mathrm{S}$.)
The simulations reproduce well the separation into two manifolds, each in turn further split in two. This supports the speculation that the zero-field splitting of the upper manifold reflects interdot tunneling, especially since that manifold was observed to decrease in frequency with more negative interdot detuning~\cite{Laird2013}. It also reproduces the branch of the upper manifold that decreases in frequency with increasing perpendicular field [Fig.~\ref{spectrum2dots}(b,h)]. However, several features of the data are not reproduced. The slopes of the resonant lines in the simulations, corresponding to the $g$-factors of the transitions, do not closely match those measured; they are too large for parallel field and too small for perpendicular. Instead, they are quite close to the single-particle $g$-factors used as inputs in the simulation, which in turn are taken from high-field d.c.\ transport in the same device~\cite{Laird2013, footnote1}. Another feature not reproduced is the zero-field splitting in the lower manifold, which is unexpected because triplet states are degenerate at zero field.

\subsection{Two coupled quantum dots and an impurity spin}

To investigate this zero-field splitting more carefully, we included in the simulation an impurity spin, coupled to one of the qubits but not participating in transport. Such an impurity has been invoked to explain transport resonances in nanotube Pauli blockade, and might arise from a paramagnetic impurity or adsorbed molecule~\cite{Chorley2011}. We suppose that the impurity spin is coupled with the electron qubit via an isotropic Heisenberg interaction~\cite{Chorley2011}
\begin{eqnarray}
H_\text{eI} &=& \frac{\JI}{4} (\sigma_\text{e}^z\sigma_\text{I}^z + \sigma_\text{e}^x\sigma_\text{I}^x + \sigma_\text{e}^y\sigma_\text{I}^y),
\end{eqnarray}
where $\{\sigma_\text{I}^{i=x,y,z}\}$ are Pauli operators of the impurity spin, and with the magnetic field via Zeeman coupling with $g$-factor $g_\mathrm{I}$.
The total Hamiltonian of the system is
\begin{equation}
H_\text{DDI} = H_\text{DD} + H_\text{eI} + \frac{1}{2} \muB g_\mathrm{I} \mathbf{B} \cdot \bfsigma_\text{I}.
\label{eq:Htotal}
\end{equation}

\begin{figure}[tbp]
\includegraphics[width=0.95\linewidth]{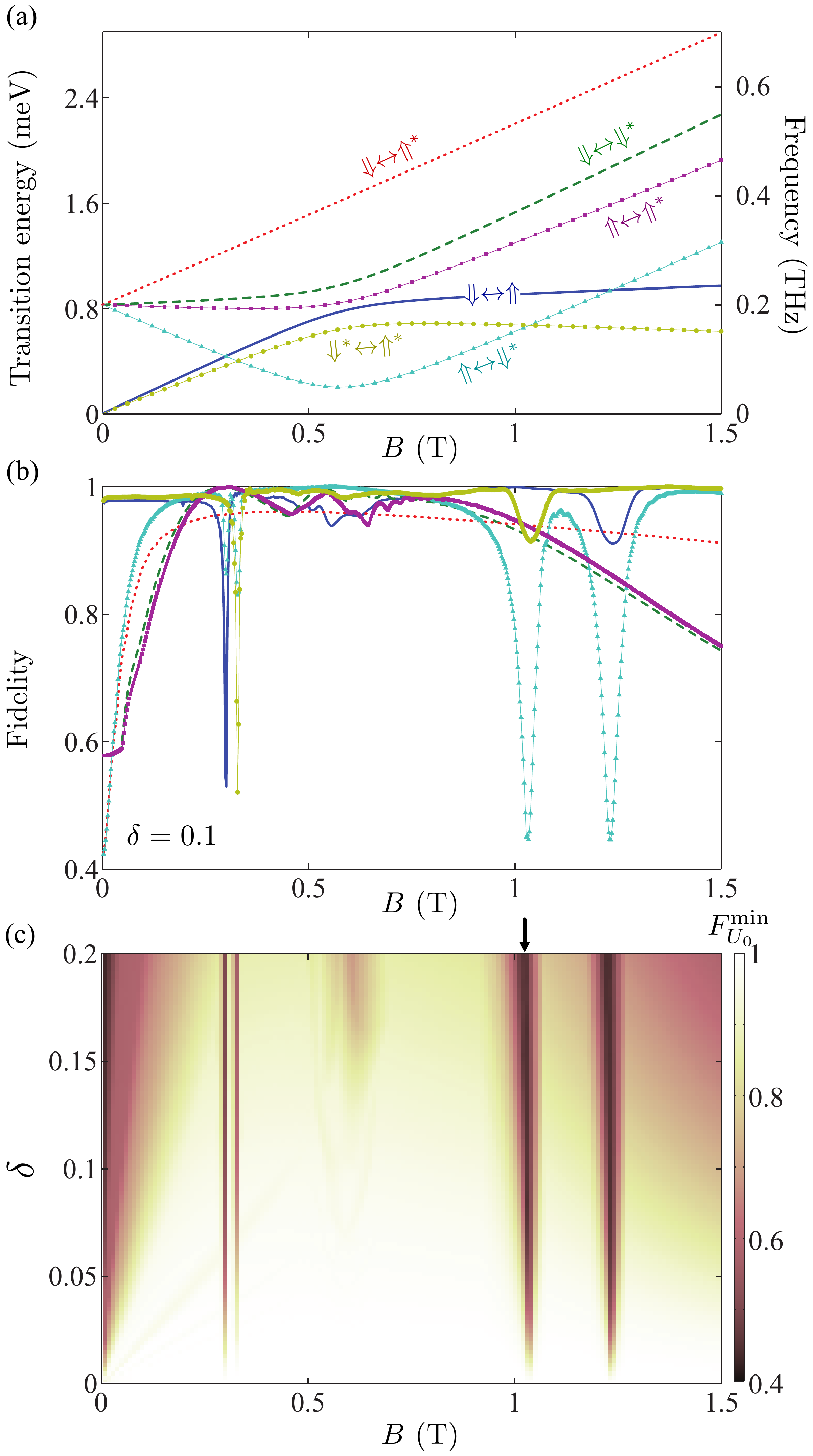}
\caption{
(a) Energy differences between valley-spin states as a function of parallel magnetic field (parameters as in Fig.~\ref{setup}(b)). (b) Fidelity of qubit rotations on each of the transitions in (a). Equal driving via all channels is assumed, characterized by a single parameter $\delta \equiv \delta\theta = \delta_{KK'} = \delta\varphi$. For this plot, $\delta=0.1$. (c) Minimum fidelity among the six transitions, plotted as a function of $B$ and $\delta$. For all values of $B$, fidelity is highest for weak driving, where individual transitions can be precisely addressed. The fidelity drops at values of $B$ where transitions in (a) become near-degenerate.
}
\label{Fid}
\end{figure}

The resulting energy levels are shown in Fig.~\ref{spectrumdefect}(a) as a function of parallel field, and other orientations are shown in Appendix Fig.~\ref{spectrumApp}. There are now two zero-field splittings, $\Delta_\pm = [\sqrt{J_0^2-J_0\JI+\JI^2} \pm (J_0+\JI)]/2$, reflecting the two exchange-like interactions in the model. In the limit of weak impurity coupling, $\Delta_+\rightarrow J_0$ and $\Delta_-\rightarrow \JI$. This can be seen in Fig.~\ref{spectrumdefect}(a); the lower manifold is split in two by the Zeeman coupling of the impurity, while the upper manifold is split in a more complicated pattern due to coupling between the $\tilde{AS}$ states of two dots and the impurity spin.

The resulting EDSR spectra are shown in Fig.~\ref{spectrumdefect}(b-d), calculated in the same way as in Fig.~\ref{spectrum2dots}. Parameters $\Delta_+$ and $\Delta_-$ are chosen to match the zero-field splittings measured in~\cite{Laird2013}. Including the impurity spin does indeed reproduce the observed zero-field splitting of the lower manifold. However, instead of the sharp kinks in the resonance lines observed for parallel field orientation~\cite{Laird2013}, the simulation gives more rounded minima. Also, the pattern of resonances simulated for the upper manifold does not agree even qualitatively with observations. We conclude that coupling to an impurity spin is not the dominant reason for the disagreement between Fig.~\ref{spectrum2dots} and experiment.

\section{EDSR of the full valley-spin Hilbert space}

The previous sections discussed manipulation within the two-state valley-spin subspace. In this section, we show that transitions within the full four-state valley-spin manifold of a single electron can be driven, provided that the applied magnetic field is well chosen and the distribution of disorder is smooth but inhomogeneous. This gives access to an effective two-qubit Hilbert space using a single electron. We evaluate the fidelity of $\pi$ pulses between all pairs of states, and show that with appropriately chosen magnetic field, high-fidelity operation can be achieved.

We take the energy quantum number $\{e,g\}$ defined in Section \ref{EDSR} to encode one qubit, and the Kramers quantum number $\{0,1\}$ to encode the other. For precise addressing, it is necessary that the six transitions between the states $\{\ket{\Uparrow}, \ket{\Downarrow}, \ket{\Uparrow^*}, \ket{\Downarrow^*}\}$ [see Fig.~\ref{setup} (b) and Appendix~\ref{AppendixA}] be separated in frequency. This is achieved with a magnetic field directed parallel to the nanotube at the quantum dot location. The transition frequencies are shown in Fig.~\ref{Fid}(a). At low field, several transitions are close together due to Kramers degeneracy and so the operating regime of qubit manipulation exceeds the regime where the perturbation theory outlined in Section~\ref{EDSR} is valid. Therefore we investigate EDSR of the full valley-spin Hilbert space by numerical integration of the time evolution governed by Eq.~(\ref{eq:Hz}).

At $\mathbf{B}=0$, electric driving combines with inhomogeneous disorder to flip only the energy qubit, but it combines with spin-orbit coupling to flip both the Kramers qubit and the energy qubit simultaneously~(see Appendix~\ref{AppendixB}). However, at finite field, spin-orbit coupling can flip the Kramers qubit separately from the energy qubit, which allows full control of the four-state Hilbert space. The matrix elements of the six transitions are
\begin{eqnarray}
V'_{\Uparrow, \Downarrow} = V'_{\Uparrow^*, \Downarrow^*} &=& \absLR{ \frac{\delta\theta}{2} \sin\delta \chi } \DeltaSO, \\
V'_{\Downarrow^*, \Downarrow} = V'_{\Uparrow^*, \Uparrow} &=& \absLR{ \frac{\delta\theta}{2} \cos\delta \chi } \DeltaSO, \\
V'_{\Uparrow^*, \Downarrow} &=& \absLR{ \frac{\delta_{KK'}}{2} \cos \chi_\downarrow +i\frac{\delta\varphi}{2} } \DeltaKK, \\
V'_{\Downarrow^*, \Uparrow} &=& \absLR{ \frac{\delta_{KK'}}{2} \cos \chi_\uparrow +i\frac{\delta\varphi}{2} } \DeltaKK,
\end{eqnarray}
where $V'_{a,b} \equiv \abs{\bra{a}V'\ket{b}}$, $\delta \chi \equiv \abs{\chi_\uparrow -\chi_\downarrow}/2$, and $\chi_\uparrow$ and $\chi_\downarrow$ are defined in Appendix~\ref{AppendixA}. When the magnetic field is weak ($\gorb\muB B_z \ll \DeltaSO$), $\delta \chi \sim 0$, transitions $\Downarrow \leftrightarrow \Uparrow$ and $\Downarrow^* \leftrightarrow \Uparrow^*$ will be inefficient; similarly, when the magnetic field is too strong ($\gorb\muB B_z \gg \DeltaSO$), $\delta \chi \sim \pi/2$, transitions $\Downarrow \leftrightarrow \Downarrow^*$ and $\Uparrow \leftrightarrow \Uparrow^*$ will be inefficient. Therefore, one condition for efficient EDSR between all states is a moderate magnetic field ($\gorb\muB B_z \sim \DeltaSO$) providing $\delta \chi \sim \pi/4$.

To identify the optimal operating condition for this two-qubit control, we examine the fidelities of six possible gate operations. The fidelity of a gate is defined as 
\begin{equation}
F_{U_0}=\int d\psi \abs{\bra{\psi}U_0^\dag U\ket{\psi}}^2,
\end{equation}
where $U_0$ and $U$ are unitary operators corresponding respectively to the target gate (a $\pi$ pulse) and to the actual evolution. Here the integration represents an average over the four-state Hilbert space~\cite{Nielsen2002}. The figure of merit for four-state control is defined as the minimum fidelity $F^\mathrm{min}_{U_0}$ among the six possible choices of $U_0$. Assuming that the angle of the nanotube and the valley mixing induced by disorder change linearly with $\delta z$ (i.e.~the bend radius is much larger than $\delta z_\mathrm{A}$ and disorder varies weakly in the range of the displacement) and the displacement oscillates harmonically, we numerically obtain the minimum fidelity under different magnetic fields and displacement amplitudes as shown in Fig.~\ref{Fid}. A larger displacement amplitude leads to faster driving, but usually to a correspondingly reduced fidelity. This is because a~$\pi$ burst addressed to one transition contains spectral components that address other transitions, and this spectral leakage becomes more severe for faster driving. Superimposed on this general behaviour are dips at field strengths where two or more transitions approach degeneracy. For example, at the field indicated by the arrow in Fig.~\ref{Fid}(c), the transitions $\Uparrow \leftrightarrow \Downarrow^*$ and $\Downarrow^* \leftrightarrow \Uparrow^*$ approach degeneracy, leading to reduced fidelity of both operations [Fig.~\ref{Fid}(b)] and correspondingly in $F^\mathrm{min}_{U_0}$. Weaker features running diagonally in~Fig.~\ref{Fid}(d) arise when detuned Rabi oscillations at a non-addressed transition (with angular frequency $\sqrt{\Delta \omega_\mathrm{d}^2+\Omega_d^2}$, where $\Delta\omega_\mathrm{d}$ and $\Omega_\mathrm{d}$ are the detuning and Rabi frequencies of the detuned transition) execute an integer number of $\pi$ rotations. This simulation does not use shaped pulses, which could improve the control fidelity~\cite{Gambetta2011}.

For optimum four-state control, the magnetic field should be set away from one of the fidelity minima. Taking $\DeltaSO=0.8\text{ meV}$, $\DeltaKK=0.2\text{ meV}$, $\gs=2$, and $\gorb=24$, Fig.~\ref{Fid}(d) suggests optimum values $B \sim 0.4\text{ T}$ or $B \sim 0.8\text{ T}$. At these values, the operation time is less than $1\text{ ns}$ when $\delta\theta = \delta_{KK'} = \delta\varphi =0.02$ for both field strengths. If the same coherence time measured in~\cite{Laird2013} applies to all transitions, this means $\sim 60$ coherent operations are possible. As seen from Fig.~\ref{Fid}(a), this scheme requires substantial driving frequencies, at least $\DeltaSO/h \sim 200\text{ GHz}$, to access all states.

\section{Conclusions}

We have presented a comprehensive theoretical study of bent and disordered nanotube qubit devices. Our analysis followed the pioneering treatment in Ref.~\cite{Flensberg2010}, but retained all terms to second order in the quantum dot displacement and in the magnetic field. We found that fully accounting for these terms alters the dependence of the EDSR Rabi frequency on magnetic field orientation (Fig.~\ref{RFbend}). We then extended the core model to include both interdot coupling, and coupling to an external impurity. The model's predictions showed qualitative agreement with previously unexplained experimental data~\cite{Laird2013}, in particular the zero field splitting of certain transition frequencies (Fig.~\ref{spectrumdefect}). Finally, we extended the model to include complete two-qubit manipulation in the four-dimensional spin-valley Hilbert space.

There are several aspects of the prior experimental data which are not explained within our model. For example, the inferred $g$-factors (taken from the slope of resonance frequencies with magnetic field) do vary with the severity of disorder in the tube; however this variation is not sufficient to explain the $g$-factors seen experimentally (see Fig.~\ref{spectrum2dots} and caption). Moreover the detailed structure of certain of our predicted curves does differ from published data. Thus there is scope for this model to be taken further, perhaps after additional experiments.

\begin{acknowledgments}
This work was supported by EPSRC platform Grant `Molecular Quantum Devices' (EP/J015067/1), Templeton World Charity Foundation, and the Royal Academy of Engineering. We acknowledge~A.~P\'{a}lyi for discussions.
\end{acknowledgments}

\appendix

\section{Eigenstates and eigenenergies}
\label{AppendixA}

With a magnetic field along the $z$-direction, the quantum dot is described by the Hamiltonian $H_B \equiv H(z_0)|_{\mathbf{B}=B_z\kk}$. As in the main text, we use $\ket{K'}$ and $\ket{K}$ ($\ket{\uparrow}$ and $\ket{\downarrow}$) to denote the valley (spin) eigenstates of $\tau_3$ ($\sigma_z$) with eigenvalues $+1$ and $-1$, respectively, and set that $\nn(z_0)=\kk$ and $\varphi(z_0)=0$ for simplification. The four eigenstates are
\begin{eqnarray}
\ket{\Uparrow} &=& \cos \frac{\chi_\uparrow}{2} \ket{K'\uparrow} + \sin \frac{\chi_\uparrow}{2} \ket{K\uparrow}, \\
\ket{\Downarrow} &=& \sin \frac{\chi_\downarrow}{2} \ket{K'\downarrow} + \cos \frac{\chi_\downarrow}{2} \ket{K\downarrow}, \\
\ket{\Downarrow^*} &=& \sin \frac{\chi_\uparrow}{2} \ket{K'\uparrow} - \cos \frac{\chi_\uparrow}{2} \ket{K\uparrow}, \\
\ket{\Uparrow^*} &=& \cos \frac{\chi_\downarrow}{2} \ket{K'\downarrow} - \sin \frac{\chi_\downarrow}{2} \ket{K\downarrow},
\end{eqnarray}
and the corresponding eigenenergies are
\begin{eqnarray}
E_{\Uparrow} &=& \frac{1}{2} \gs\muB B_z -\frac{1}{2}\Delta E_{\uparrow}, \\
E_{\Downarrow} &=& -\frac{1}{2} \gs\muB B_z -\frac{1}{2}\Delta E_{\downarrow}, \\
E_{\Downarrow^*} &=& \frac{1}{2} \gs\muB B_z +\frac{1}{2}\Delta E_{\uparrow}, \\
E_{\Uparrow^*} &=& -\frac{1}{2} \gs\muB B_z +\frac{1}{2}\Delta E_{\downarrow}.
\end{eqnarray}
Here we define
\begin{eqnarray}
\cos \chi_\uparrow &=& \frac{\DeltaSO -\gorb\muB B_z}{\Delta E_{\uparrow}}, \\
\sin \chi_\uparrow &=& \frac{\DeltaKK(z_0)}{\Delta E_{\uparrow}}, \\
\cos \chi_\downarrow &=& \frac{\DeltaSO +\gorb\muB B_z}{\Delta E_{\downarrow}}, \\
\sin \chi_\downarrow &=& \frac{\DeltaKK(z_0)}{\Delta E_{\downarrow}},
\end{eqnarray}
and
\begin{eqnarray}
\Delta E_{\uparrow} &=& \sqrt{(\DeltaSO-\gorb\muB B_z)^2+\DeltaKK^2(z_0)}, \\
\Delta E_{\downarrow} &=& \sqrt{(\DeltaSO+\gorb\muB B_z)^2+\DeltaKK^2(z_0)}.
\end{eqnarray}

At $\mathbf{B}=0$, we have $\Delta E_{\uparrow} = \Delta E_{\downarrow} = \Delta E_0$ and $\chi_\uparrow = \chi_\downarrow = \chi$. Therefore, states $\ket{g 0} = \ket{\Uparrow}$ and $\ket{g 1} = \ket{\Downarrow}$ form the lower doublet with energy $E_{\Uparrow} = E_{\Downarrow} = -\Delta E_0/2$, and states $\ket{e 0} = \ket{\Downarrow^*}$ and $\ket{e 1} = \ket{\Uparrow^*}$ form the higher doublet with energy $E_{\Downarrow^*} = E_{\Uparrow^*} = \Delta E_0/2$.

\section{Perturbation theory}
\label{AppendixB}

The Pauli operators $\{\sigma_i, \tau_i\}$ of the spin states and valley states can be expressed in terms of Pauli operators $\{\sigma_\text{E}, \sigma_\text{K}\}$ of the energy qubit and the valley-spin qubit as follows:
\begin{eqnarray}
\tau_1&=&\sin \chi\sigma_\text{E}^z - \cos \chi\sigma_\text{E}^x\sigma_\text{K}^z, \\
\tau_2&=&-\sigma_\text{E}^y, \\
\tau_3&=&\sin \chi\sigma_\text{E}^x + \cos \chi\sigma_\text{E}^z\sigma_\text{K}^z,
\end{eqnarray}
\begin{eqnarray}
\sigma_x&=&\sin \chi\sigma_\text{K}^x - \cos \chi\sigma_\text{E}^y\sigma_\text{K}^y, \\
\sigma_y&=&\sin \chi\sigma_\text{K}^y + \cos \chi\sigma_\text{E}^y\sigma_\text{K}^x, \\
\sigma_z&=&\sigma_\text{K}^z.
\end{eqnarray}
With these expressions, the perturbation $V^{(1)}$ [see Eqs.~(\ref{eq:Vz})~and~(\ref{eq:VB})] can be written
\begin{eqnarray}
V^{(1)}&=&-\frac{1}{2} \delta\theta \DeltaSO \sigma_\text{E}^x\sigma_\text{K}^x \notag \\
&&-\frac{1}{2} \delta_{KK'} \DeltaKK (\sin \chi\sigma_\text{E}^z - \cos \chi\sigma_\text{E}^x\sigma_\text{K}^z) \notag \\
&&+\frac{1}{2} \delta\varphi \DeltaKK \sigma_\text{E}^y \notag \\
&&+\frac{1}{2} \gs\muB [
B_x(\sin \chi\sigma_\text{K}^x - \cos \chi\sigma_\text{E}^y\sigma_\text{K}^y) \notag \\
&&\hspace{1.4cm}+B_y(\sin \chi\sigma_\text{K}^y + \cos \chi\sigma_\text{E}^y\sigma_\text{K}^x) \notag \\
&&\hspace{1.4cm}+B_z\sigma_\text{K}^z ] \notag \\
&&+\frac{1}{2} \gorb\muB B_z (\cos \chi\sigma_\text{E}^z\sigma_\text{K}^z + \sin \chi\sigma_\text{E}^x).
\end{eqnarray}
From this equation, one sees that the effect of spin-orbit coupling (first term) is to flip the energy qubit and the valley-spin qubit together, while the effect of valley mixing is to flip the energy qubit alone and to shift the energy of an entire doublet.

Since all terms in the perturbation $V^{(1)}$ are products of scalars and Pauli operators, a term $A$ and a term~$B$ in $V^{(1)}$ contribute to $H_\text{eff}^{(2)}$ either a nontrivial term $-\sigma_\text{E}^z\{A,B\}/\Delta E_0$ or a trivial term proportional to $\sigma_\text{E}^z$ up to a scalar factor. The contribution is nontrivial iff both $A$ and $B$ contain either $\sigma_\text{E}^x$ or $\sigma_\text{E}^y$, $A\neq B$, and $[A,B] = 0$.

\section{Two-dot coupling}
\label{AppendixC}

\begin{figure}[tbp]
\includegraphics[width=1\linewidth]{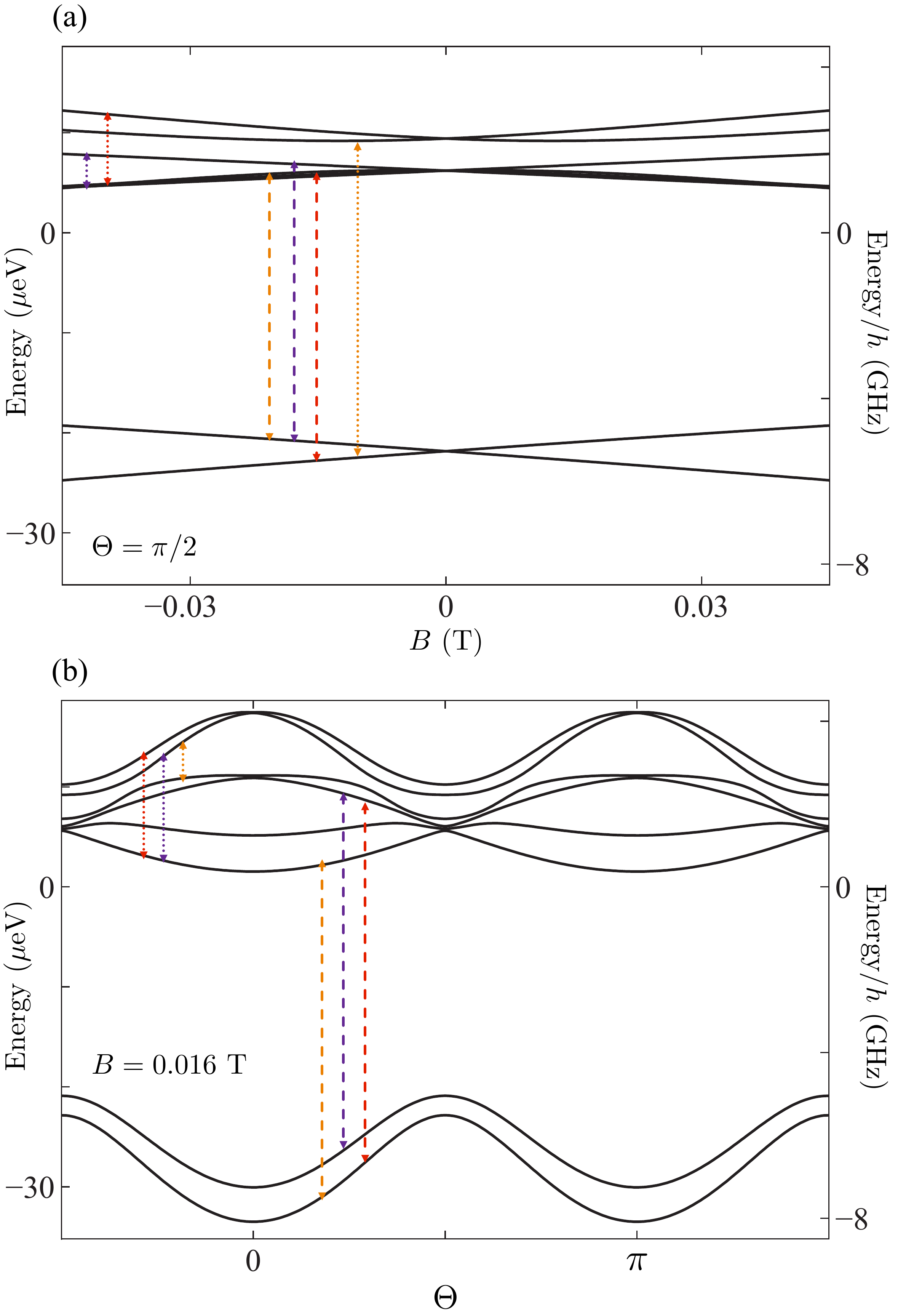}
\caption{
Energy levels for two coupled valley-spin qubits and an impurity spin, for the same parameters as Fig.~\ref{spectrumdefect}. Panel (a) is calculated as a function of $B$ perpendicular to the nanotube, panel (b) as a function of field angle for $B=0.016$~T. The transitions highlighted in Fig.~\ref{spectrumdefect}(c,d) are indicated.
}
\label{spectrumApp}
\end{figure}

We consider a single-electron quantum dot and a single-hole quantum dot as in Fig.~\ref{setup}~(a). Similar discussion can be found in Ref.~\cite{Laird2013}. Valley-spin qubit states of the electron quantum dot and the hole quantum dot are $\{\ket{g 0}_\text{e},\ket{g 1}_\text{e}\}$ and $\{\ket{e 0}_\text{h},\ket{e 1}_\text{h}\}$, respectively. Here we define $\{\ket{e 0}_\text{h},\ket{e 1}_\text{h}\}$ to mean a filled lower doublet with a third electron occupying the corresponding state in the upper doublet. Then
\begin{eqnarray}
\ket{g 0}_\text{e} &=& \cos \frac{\chi}{2} \ket{K'\uparrow}_\text{e} + \sin \frac{\chi}{2} \ket{K\uparrow}_\text{e},\\
\ket{g 1}_\text{e} &=& \sin \frac{\chi}{2} \ket{K'\downarrow}_\text{e} + \cos \frac{\chi}{2} \ket{K\downarrow}_\text{e}
\end{eqnarray}
while
\begin{eqnarray}
\ket{e 0}_\text{h} &=& \sin \frac{\chi_\text{h}}{2} \ket{K'\uparrow}_\text{h} - e^{i\varphi_\text{h}} \cos \frac{\chi_\text{h}}{2} \ket{K\uparrow}_\text{h}, \\
\ket{e 1}_\text{h} &=& \cos \frac{\chi_\text{h}}{2} \ket{K'\downarrow}_\text{h} - e^{i\varphi_\text{h}}\sin \frac{\chi_\text{h}}{2} \ket{K\downarrow}_\text{h},
\end{eqnarray}
where $\chi_\text{h}$, similar to $\chi$, is defined by $\cos \chi_\text{h} = \DeltaSO/\Delta E_\text{h}$, $\sin \chi_\text{h} = \DeltaKKh/\Delta E_\text{h}$, and the energy gap is $\Delta E_\text{h}=\sqrt{\DeltaSO^2+\DeltaKKh^2}$. The tunneling matrix elements are proportional to the overlaps
\begin{eqnarray}
{_\text{e}}\braket{g 0}{e 0}_\text{h} &=& \cos\frac{\chi}{2}\sin\frac{\chi_\text{h}}{2} - e^{i\varphi_\text{h}} \sin\frac{\chi}{2}\cos\frac{\chi_\text{h}}{2}, \label{ol0} \\
{_\text{e}}\braket{g 1}{e 1}_\text{h} &=& \sin\frac{\chi}{2}\cos\frac{\chi_\text{h}}{2} - e^{i\varphi_\text{h}}\cos\frac{\chi}{2}\sin\frac{\chi_\text{h}}{2}, \label{ol1}
\end{eqnarray}
and ${_\text{e}}\braket{g 0}{e 1}_\text{h} = {_\text{e}}\braket{g 1}{e 0}_\text{h} = 0$.

If the hole states are regauged ($\ket{e0}_\text{h} \rightarrow e^{i\alpha} \ket{e0}_\text{h}$, $\ket{e1}_\text{h} \rightarrow \ket{e1}_\text{h}$) so that the overlaps (\ref{ol0}-\ref{ol1}) become identical, this situation becomes isomorphic to that of two spins in a conventional double dot. Here the phase factor $\alpha$ is equal to the phase difference between the two overlaps Eqs.~(\ref{ol0}-\ref{ol1}):
\begin{eqnarray}
\alpha = \pi -\varphi_\text{h} +2\text{arg}(_\text{e}\braket{g 0}{e 0}_\text{h}).
\end{eqnarray}
The tunnel coupling  opens up an energy difference $J_0$ between longitudinally symmetric and antisymmetric states \cite{Loss1998}, equivalent to an exchange Hamiltonian
\begin{eqnarray}
H_\text{eh}' = - J_0 \ketbra{S}{S},
\label{eq:Hehprime}
\end{eqnarray}
where the symmetric state is
\begin{eqnarray}
\ket{S} = \frac{1}{\sqrt{2}}(\ket{g 0}_\text{e}\ket{e 1}_\text{h} - e^{-i\alpha}\ket{g 1}_\text{e}\ket{e 0}_\text{h}).
\end{eqnarray}

By rewriting Eq.~(\ref{eq:Hehprime}) in terms of the valley-spin Pauli operators, the coupling Hamiltonian Eq.~(\ref{H2dots}) is obtained, up to an insignificant additive constant.


\begin{thebibliography}{9}
\bibitem{footnote1} High-field magnetic transport data are taken to yield single-particle $g$-factors because the splitting of states with different magnetic moments should overwhelm any exchange-like coupling.
\end{thebibliography}
\end{document}